\documentclass{article}
\usepackage{amssymb}

\usepackage{amsmath}

\input{tcilatex}

\begin{document}

\title{The Real Scalar Field in Schwarzchild-de Sitter Spacetime\thanks{%
Supported by the National Natural Science Foundation of China under Grant
No. 10275008.}}
\author{Jianxiang Tian\thanks{%
Email address: lanmanhuayu@yahoo.com.cn}, Yuanxing Gui \thanks{%
Email address: guiyx@dlut.edu.cn. Telephone number: 86-0411-4706203} \and %
Guanghai Guo, Yan Lv, Suhong Zhang and Wei Wang \\
\\
\textit{Department of Physics, Dalian University of Technology }\\
\textit{\ Dalian 116024, P.R.China}}
\maketitle

\begin{abstract}
In this paper, the real scalar field equation in Schwarzschild-de Sitter
spacetime is solved numerically with high precision. A method called
`polynomial' approximation is introduced to derive the relation between the
tortoise coordinate $x$ and the radius $r$. This method is different from
the `tangent' approximation [1] and leads to more accurate result. The
Nariai black hole is then discussed in details. We find that the wave
function is harmonic only near the horizons as I. Brevik and B. Simonsen $%
[1] $ found. Howerver the wave function is not harmonic in the region of the
potential peak, with amplitude increasing instead. Furthermore, we also find
that, when cosmological constant decreases, the potential peak increases,
and the maximum wave amplitude increases.
\end{abstract}

\section{\protect\bigskip Introduction}

I. Brevik and B. Simonsen $[1]$ solved the real scalar field equation in
Schwarzchild-de Sitter spacetime numerically. They approximated the tortoise
coordinate $x=x(r)$ by the `tangent' function $\overset{\symbol{126}}{r}=%
\overset{\symbol{126}}{r}(x)$ and replaced the potential function $v=v(r)$
by $v=v(\overset{\symbol{126}}{r})=v(x)$. The tortoise coordinate $x$\ was
introduced to simply the real scalar field equation. Then they solved the
equation by Matlab software, and found that the solution in SDS system is
close to that of a harmonic wave $[1]$.

In the present paper, we continue the previous work and investigate the
solution of the real scalar field equation in Schwarzchild-de Sitter
spacetime. In section 2, we will introduce a new method, referred to as the
`polynomial' approximation, to get the relation between the tortoise
coordinate $x$ and the radius $r$. This approximation is different from the
`tangent' approximation introduced by I. Brevik and B. Simonsen $[1]$, and
approximates the curve of the tortoise coordinate $x$ versus the radius $r$
more accurately. Then we solve the real scalar field equation by Matlab
software. The solution illustrated in figures demonstrates that the solution
of the real scalar field equation in Schwarzschild-de Sitter spacetime is
not harmonic globally in contrast to the result from the work in Ref.(1).

We adopt the signature ($-+++$), put$\ \hbar ,\ c,\ $and $G$ (Newton's
gravitational constant) equal to unity, and follow the conventions of Misner 
$et$ $al.[2]$

\section{Global Solution of the Real Scalar Field Equation in Schwarzschild-
de Sitter spacetime}

\subsection{\protect\bigskip Horizons}

With our conventions Einstein's equations read

\begin{equation}
R_{\mu \nu }-\frac{1}{2}Rg_{\mu \nu }+\Lambda g_{\mu \nu }=8\pi T_{\mu \nu }.
\end{equation}%
For a spherically symmetric system the line element takes the form

\begin{equation}
ds^{2}=-f(r)dt^{2}+\frac{1}{f(r)}dr^{2}+r^{2}d\theta ^{2}+r^{2}\sin
^{2}\theta d\varphi ^{2}.
\end{equation}%
In our case

\begin{equation}
f(r)=1-\frac{2M}{r}-\frac{\Lambda r^{2}}{3},
\end{equation}%
where $M$ is the mass of the black hole. I. Brevik and B. Simonsen[1] have
discussed that when $\Lambda M^{2}<\frac{1}{9}$, equation $f(r)=0$ has three
real solutions. They are event horizon $r_{e},\ $cosmological horizon $r_{c}$
and$\ $a negative solution $r_{o}$, with $r_{e}>0$, $r_{c}>0$, $%
r_{o}=-(r_{c}+r_{e}).\ $The choice $\Lambda M^{2}=0.11$ leads to $%
r_{e}=2.8391M$ and $r_{c}=3.1878M$.

\subsection{\protect\bigskip The Wave Function}

\bigskip The real scalar field equaion in this case can be written as%
\begin{equation}
\square \Phi =-\frac{\Phi _{,tt}}{f(r)}+\frac{1}{r^{2}}\left[ r^{2}f(r)\Phi
_{,r}\right] _{,r}+\frac{1}{r^{2}\sin \theta }\left[ (\sin \theta \Phi
_{,\theta })_{,\theta }+\frac{\Phi _{,\varphi \varphi }}{\sin \theta }\right]
=0.
\end{equation}%
Let

\begin{equation}
\Phi =\frac{1}{r}\Psi _{\omega l}(r)e^{-i\omega t}Y_{lm}(\theta ,\varphi ),
\end{equation}%
where $Y_{lm}$ is the usual spherical harmonic. We have

\begin{equation}
\left[ -f(r)\frac{d}{dr}\left( f(r)\frac{d}{dr}\right) +v(r)\right] \Psi
_{\omega l}(r)=\omega ^{2}\Psi _{\omega l}(r),
\end{equation}%
where $v(r)$ is the potential

\begin{equation}
v(r)=f(r)\left[ \frac{f^{\prime }(r)}{r}+\frac{l(l+1)}{r^{2}}\right] =\left(
1-\frac{2M}{r}-\frac{\Lambda r^{2}}{3}\right) \left( \frac{2M}{r^{3}}-\frac{%
2\Lambda }{3}+\frac{2}{r^{2}}\right) .
\end{equation}%
$l$ is taken as unity here. We now introduce the tortoise coordinate $x$ by
the equation [3]

\begin{equation}
x=\frac{1}{2M}\int \frac{dr}{f(r)}.
\end{equation}%
This quantity is conveniently expressed in terms of the surface gravities $%
\kappa _{i},$ defined by [4]

\begin{equation}
\kappa _{i}=\frac{1}{2}\left| \frac{df}{dr}\right| _{r=r_{i}}.
\end{equation}%
We get

\begin{equation}
\kappa _{e}=\frac{\left( r_{c}-r_{e}\right) \left( r_{e}-r_{o}\right) }{%
6r_{e}}\Lambda ,
\end{equation}

\begin{equation}
\kappa _{c}=\frac{\left( r_{c}-r_{e}\right) \left( r_{c}-r_{o}\right) }{%
6r_{c}}\Lambda ,
\end{equation}

\begin{equation}
\kappa _{o}=\frac{\left( r_{o}-r_{e}\right) \left( r_{c}-r_{o}\right) }{%
6r_{o}}\Lambda ,
\end{equation}%
The tortoise coordinate can now be written as

\begin{equation}
x=\frac{1}{2M}\left[ \frac{1}{2\kappa _{e}}\ln \left( \frac{r}{r_{e}}%
-1\right) -\frac{1}{2\kappa _{c}}\ln \left( 1-\frac{r}{r_{c}}\right) +\frac{1%
}{2\kappa _{o}}\ln \left( 1-\frac{r}{r_{o}}\right) \right] .
\end{equation}

\bigskip From Eq. (8), we get

\begin{equation}
dx/dr=\frac{1}{2f(r)}.
\end{equation}%
So Eq. (6) is rewritten as 
\begin{equation}
\left[ -\frac{d^{2}}{dx^{2}}+4M^{2}v\right] \Psi _{\omega l}(x)=4M^{2}\omega
^{2}\Psi _{\omega l}(x),
\end{equation}%
where $v$ is now a function of $x.$%
\begin{equation}
v=v(x).
\end{equation}

\subsection{Boundary Conditions and Results}

To proceed further, we need the inversion $r=r(x)$ of the tortoise
coordinate $x=x(r)$ to get $v=v(r)=v(r(x))=v(x)$. But it is difficult to
invert Eq. (13). In order to get the function $r=r(x)$, we turn to the
numerical way. For Eq. (13) is a one to one mapping function, we use the
command `polyfit' in Matlab software to get a new function $%
y=y(x)=\sum_{i=0}^{n}a_{i}x^{i}$, $n\in N$, $i\in N$, to approximate
tortoise coordinate $r=r(x)$. In different interval of the radius $r$, the
function $y$ has different form. In the same interval, $n$ is determined by
the demand of our approximation accuracy. The bigger $n$ is, the higher the
accuracy is. Now we put $M=1$, $\Lambda =0.11$ and consider Nariai black
hole $[1].$ Thus the horizons are $r_{e}=2.8391$ and $r_{c}=3.1878.$ In the
interval $r\in \lbrack 2.83908,3.18775]$, we have

\begin{equation}
y=\sum_{i=0}^{20}a_{i}x^{i}.
\end{equation}%
$\{a_{i}\}$ are displayed in $Table$ $1.$

The curve of $y$ versus the tortoise coordinate $x$ and the curve of the
radius $r$ versus the tortoise coordinate $x$ are illustrated in Figure 1,
from which we know that the two curves overlap with each other in the
interval $r\in \lbrack 2.83908,3.18775]$. In figures, point A denotes the
position at which the potential peak is. Now we see that \ the function $y$
may be better as an approximation to Eq. (13) than the tangent function [1] $%
\overset{\symbol{126}}{r}=15\tan [b(r-d)+5]$, with $b=2.7/(r_{c}-r_{e})$, $%
d=(r_{c}+r_{e})/2$. The curve of $\overset{\symbol{126}}{r}$ versus the
tortoise coordinate $x$ and the curve of the radius $r$ versus the radius $x$
were illustrated in Ref. (1).

Now inserting $y=y(x)$ into form (7) instead of $r=r(x)$, we obtain the
potential $v$ as a function of $x$ 
\begin{equation}
v=v(x).
\end{equation}%
The curve of $v$ versus $x$ from equation (13) and the curve of $v$ versus $%
x $ from the approximation (17) is illustrated in Figure 2. The curve of $v$
versus $y$ and the curve of $v$ versus the radius $r$ are illustrated in
Figure 3. Figure 2 and Figure 3 show that our approximation is good.

We set $v=0$ near the horizons. Thus Eq. (15) becomes

\begin{equation}
\left[ -\frac{d^{2}}{dx^{2}}\right] \Psi _{\omega l}(x)=4\Psi _{\omega l}(x),
\end{equation}%
which has the following solution%
\begin{equation}
\Psi _{\omega l}(x)=\cos (2x).
\end{equation}%
Therefore, we have [1]%
\begin{equation}
\Psi _{\omega l}(x=-100)=\Psi _{\omega l}(x=100)=\cos (200).
\end{equation}

The above numerical values are used in the boundary conditions when Eq. (15)
is solved with the Matlab software. The solution is illustrated in Figure 4
and Figure 5. These two figures show that the wave function near the
horizons is harmonic, but it isn't in the region where the potential gets
its peak value. There is a largest amplitude at $x=3.1431$ near the point $%
x=2.8582$. While the potential peak $v=7.5\times 10^{-4}$ appears at $%
x=2.8582.\ $When $r$ trends to the point $x=3.1431,$ the amplitude of the
wave increases step by step. I. Brevik and B. Simonsen [1] illustrated the
partial solution of Eq. (15) and mentioned that the solution is close to
that of a harmonic wave. Now we find that the wave function $\Psi _{\omega
l} $ is not harmonic globally.

The case when $\Lambda =0.001$ will not be discussed here again. Its result
is similar to the one when $\Lambda =0.11$. But one thing should be
mentioned. It is that, When $\Lambda $ decreases, the potential peak
increases. So does the largest amplitude of the wave.

\section{\protect\bigskip Summary}

\subsection{What We Have Done}

In this paper, we introduce a new method called `polynomial' approximation
to approximate the tortoise coordinate $x=x(r).$ And we find that the
solution of the real scalar field equation in Schwarzchild-de Sitter
spacetime is harmonic near the horizons as I. Brevik and B. Simonsen $[1]$
found, and that it isn't near the position where the potential gets its peak
value, with higher wave peaks and lower troughs appearing instead.

\subsection{About the Reflection and Transmission Coefficient}

In this paper, we did not calculate the reflection and transmission
coefficients. The reason is that there are many square potentials [1] to
describe a common potential $v=v(x)$. Each square potential gives out a
reflection coefficient and a transmission coefficient by the method adopted
in I. Brevik and B. Simonsen's paper [1]. Different square potential gives
out different reflection coefficient and transmission coefficient.

\begin{equation*}
Table1.\text{the coefficients in function }y=y(x)
\end{equation*}%
\begin{tabular}{|l|l|l|}
\hline
$a_{0}=2.9817$ & $a_{1}=6.5107\times 10^{-3}$ & $a_{2}=4.0912\times 10^{-5}$
\\ \hline
$a_{3}=-2.9913\times 10^{-6}$ & $a_{4}=-3.4895\times 10^{-8}$ & $%
a_{5}=1.6009\times 10^{-9}$ \\ \hline
$a_{6}=2.3413\times 10^{-11}$ & $a_{7}=-8.0083\times 10^{-13}$ & $%
a_{8}=-1.1964\times 10^{-14}$ \\ \hline
$a_{9}=3.3845\times 10^{-16}$ & $a_{10}=4.3110\times 10^{-18}$ & $%
a_{11}=-1.0899\times 10^{-19}$ \\ \hline
$a_{12}=-1.0120\times 10^{-21}$ & $a_{13}=2.4364\times 10^{-23}$ & $%
a_{14}=1.4031\times 10^{-25}$ \\ \hline
$a_{15}=-3.4329\times 10^{-27}$ & $a_{16}=-9.5242\times 10^{-30}$ & $%
a_{17}=2.6439\times 10^{-31}$ \\ \hline
$a_{18}=1.5652\times 10^{-34}$ & $a_{19}=-8.0402\times 10^{-36}$ & $%
a_{20}=4.3262\times 10^{-39}$ \\ \hline
\end{tabular}

\FRAME{ftbpFU}{342.5pt}{257.375pt}{0pt}{\Qcb{The curve of the tortoise
coordinate $x$ versus $r$ (full line), together with the curve of the
tortoise coordinate $x$ versus polynomial approximation $y$ (dotted). $%
\Lambda =0.11.$}}{}{Figure}{\special{language "Scientific Word";type
"GRAPHIC";maintain-aspect-ratio TRUE;display "USEDEF";valid_file "T";width
342.5pt;height 257.375pt;depth 0pt;original-width 497.9375pt;original-height
373.4375pt;cropleft "0";croptop "1";cropright "1";cropbottom
"0";tempfilename 'HB607I00.wmf';tempfile-properties "XPR";}}

\bigskip

\FRAME{ftbpFU}{342.5pt}{257.375pt}{0pt}{\Qcb{The curve of the potential $v$
versus the tortoise coordinate $x$ (dotted), together with the curve of the
potential $v$ versus $x$ from the polynomial approximation $y=y(x)$ (full
line)$.$ $\Lambda =0.11$}}{}{Figure}{\special{language "Scientific
Word";type "GRAPHIC";maintain-aspect-ratio TRUE;display "USEDEF";valid_file
"T";width 342.5pt;height 257.375pt;depth 0pt;original-width
497.9375pt;original-height 373.4375pt;cropleft "0";croptop "1";cropright
"1";cropbottom "0";tempfilename 'HB607I01.wmf';tempfile-properties "XPR";}}

\FRAME{ftbpFU}{342.5pt}{257.375pt}{0pt}{\Qcb{ The curve of the potential $v$
versus the radius $r$ (dotted), together with the curve of the potential $v$
versus $y$ (full line). $\Lambda =0.11.$}}{}{Figure}{\special{language
"Scientific Word";type "GRAPHIC";maintain-aspect-ratio TRUE;display
"USEDEF";valid_file "T";width 342.5pt;height 257.375pt;depth
0pt;original-width 497.9375pt;original-height 373.4375pt;cropleft
"0";croptop "1";cropright "1";cropbottom "0";tempfilename
'HB607I02.wmf';tempfile-properties "XPR";}}

\FRAME{ftbpFU}{342.5pt}{257.375pt}{0pt}{\Qcb{The curve of the wave function $%
\Psi _{\protect\omega l}$ versus $x$ in SDS. $\Lambda =0.11.$}}{}{Figure}{%
\special{language "Scientific Word";type "GRAPHIC";maintain-aspect-ratio
TRUE;display "USEDEF";valid_file "T";width 342.5pt;height 257.375pt;depth
0pt;original-width 497.9375pt;original-height 373.4375pt;cropleft
"0";croptop "1";cropright "1";cropbottom "0";tempfilename
'HB607I03.wmf';tempfile-properties "XPR";}}\FRAME{ftbpFU}{342.5pt}{257.375pt%
}{0pt}{\Qcb{The curve of the wave function $\Psi _{\protect\omega l}$ versus 
$r$ in SDS. $\Lambda =0.11.$}}{}{Figure}{\special{language "Scientific
Word";type "GRAPHIC";maintain-aspect-ratio TRUE;display "USEDEF";valid_file
"T";width 342.5pt;height 257.375pt;depth 0pt;original-width
497.9375pt;original-height 373.4375pt;cropleft "0";croptop "1";cropright
"1";cropbottom "0";tempfilename 'HB607J04.wmf';tempfile-properties "XPR";}}


\bigskip

\begin{thebibliography}{9}
\bibitem{1} B. Simonsen and I. Brevik, \textit{Gen}. \textit{Relativ}. 
\textit{and} \textit{Gravit. }, Vol. 33. 10. October 2001.

\bibitem{2} Misner, C.W., Thorne, K.S., and Wheeler, J.A.(1973). \textit{%
Gravitation} (W.H.Freeman and Co.,San Francisco).

\bibitem{3} Brady, P.R., Chambers, C.M., Laarakkers, W. G., and Poisson, E.
(1999). Phys. Rev. D 60, 064003.

\bibitem{4} Brady, P.R., Chambers, C.M., Krivan, W., and Laguna, P.(1997).
Phys. Rev. D 55, 7538.\bigskip
\end{thebibliography}
\end{document}